\documentclass[english,reprint, superscriptaddress, breaklinks=true, showkeys, showpacs=false, nofootinbib]{revtex4}
\usepackage[T1]{fontenc}
\usepackage[latin9]{inputenc}
\usepackage{color}
\usepackage{babel}
\usepackage{amsmath}
\usepackage{amssymb}
\usepackage{graphicx}
\usepackage[unicode=true,pdfusetitle,
 bookmarks=true,bookmarksnumbered=false,bookmarksopen=false,
 breaklinks=true,pdfborder={0 0 0},backref=false,colorlinks=true]
 {hyperref}
\usepackage{breakurl}

\makeatletter
\@ifundefined{textcolor}{}
{%
 \definecolor{BLACK}{gray}{0}
 \definecolor{WHITE}{gray}{1}
 \definecolor{RED}{rgb}{1,0,0}
 \definecolor{GREEN}{rgb}{0,1,0}
 \definecolor{BLUE}{rgb}{0,0,1}
 \definecolor{CYAN}{cmyk}{1,0,0,0}
 \definecolor{MAGENTA}{cmyk}{0,1,0,0}
 \definecolor{YELLOW}{cmyk}{0,0,1,0}
}

\hypersetup{colorlinks=true,citecolor=blue,linkcolor=cyan,urlcolor=blue,filecolor= green, breaklinks=true}
\usepackage{url}
\usepackage{breakurl}
\makeatother

\begin{document}

\title{Hilbert-Schmidt quantum coherence in multi-qudit systems}

\author{Jonas Maziero}

\email{jonas.maziero@ufsm.br}

\address{Departamento de F\'isica, Centro de Ci\^encias Naturais e Exatas, Universidade Federal de Santa Maria, Avenida Roraima 1000, 97105-900, Santa Maria, RS, Brazil}
\begin{abstract}
Using Bloch's parametrization for qudits ($d$-level quantum systems),
we write the Hilbert-Schmidt distance (HSD) between two generic $n$-qudit
states as an Euclidean distance between two vectors of observables
mean values in $\mathbb{R}^{\Pi_{s=1}^{n}d_{s}^{2}-1}$, where $d_{s}$
is the dimension for qudit $s$. Then, applying the generalized Gell
Mann's matrices to generate $SU(d_{s})$, we use that result to obtain
the Hilbert-Schmidt quantum coherence (HSC) of $n$-qudit systems.
As examples, we consider in details one-qubit, one-qutrit, two-qubit,
and two copies of one-qubit states. In this last case, the possibility
for controlling local and non-local coherences by tuning local populations
is studied and the contrasting behaviors of HSC, $l_{1}$-norm coherence,
and relative entropy of coherence in this regard are noticed. We also
investigate the decoherent dynamics of these coherence functions under
the action of qutrit dephasing and dissipation channels. At last,
we analyze the non-monotonicity of HSD under tensor products and report
the first instance of a consequence (for coherence quantification)
of this kind of property of a quantum distance measure.
\end{abstract}


\maketitle

\section{Introduction}

\label{sec:intro}

Coherent superposition states are an important aspect of quantum systems
\cite{Feynman,Wolf}, one which is believed to make possible their
use for more efficient realization of energy and information manipulation
tasks \cite{Adesso_CE,Gu,Allegra,Scholes,Uzdin,Hillery,Ma,Lewenstein-1,Pati,Fan}.
In the last few years, much attention has been given for the development,
analysis, and application of a resource theory for quantum coherence
(RTC). For a recent review, see Ref. \cite{Plenio_QC_RMP}. It is
certainly worthwhile recognizing the existence of several levels of
incoherent operations (IO) in the RTC \cite{Gour,Spekkens}. Naturally,
depending on its properties, different coherence functions may or
may not be a coherence monotone under a specific element of a given
set of IO \cite{Plenio_QC,Adesso_RoC,Adesso_RoC-1,Mintert}. This
kind of observation points for the need of further investigations
concerning the similarities and differences among the properties enjoyed
by varied quantumness functions.

Here we shall pay special attention to the geometric quantum coherence
function which is obtained using Hilbert-Schmidt's distance (HSD).
Although the possible non-contractivity under general quantum operations
of the HSD \cite{Ruskai,Schirmer,Hanggi} demands care when of its
application for quantumnesses quantification \cite{Ozawa,Piani},
it is a fact that the HSD has been a handy tool for a variety of investigations
in quantum information science \cite{Walther,Zhou,Thirring,Krammer,Brukner,Cohen,Wunsche,Sommers,Winter,Soto}.
In addition to that, recently it was shown that the HSD leads to an
observable upper bound for the robustness of coherence, which is a
full quantum coherence monotone \cite{Adesso_RoC}. Besides, because
of its friendliness for analytical calculations, the HSD can be a
useful tool for starting on quantumnesses quantification, as happened,
for instance, with quantum entanglement and quantum discord \cite{Trucks,Moor,Maziero_HSE,Brukner_HSDisc,Li,Fu_HSDisc,Azmi,Fu_MIN,Adesso_NegHSD,Hou}.

Our main goal in this article is obtaining the Hilbert-Schmidt coherence
(HSC) of multipartite qudit states and to use it to analyze the distribution
and manipulation of quantum coherence in this kind of scenario. We
shall then deal with two relevant-associated topics. We will look
at the possibility of controlling the local and/or non-local coherences
\cite{Maziero_NLQC,Byrnes,Jeong} of a system by manipulating its
local populations. And we will investigate the issue of a possible
non-monotonic behavior of quantum distance measures under tensor products
\cite{Verstraete,Maziero_NMuTP}. 

The non-monotonicity under tensor products (NMuTP) of a distance measure,
investigated recently for the trace distance \cite{Maziero_NMuTP},
is the possibility for an inversion in the dissimilarity relation
between pairs of states when we consider their copies. That is to
say, for four density matrices $\rho,\sigma,\xi,\eta$, it can happen
that
\begin{equation}
d(\rho,\sigma)>d(\xi,\eta)\mbox{ and }d(\rho\otimes\rho,\sigma\otimes\sigma)<d(\xi\otimes\xi,\eta\otimes\eta).
\end{equation}
Once there is no known analytical expression for the trace distance
coherence for general states of systems with dimension greater or
equal to three \cite{Lewenstein,Li-1,Wang}, and with the aim of unveiling
the possible implications of the NMuTP in quantum information science,
here we shall analyze it using the HSD and look for its possible implications
in coherence quantification. 

The remainder of the article is structured as follows. In Sec. \ref{sec:hsd_nqudits}
we make use of Bloch's representation to write the Hilbert-Schmidt
distance between two generic $n$-qudit states as an Euclidean distance
between the associated rescaled Bloch's vectors. In Sec. \ref{sec:hsc_nqudits}
we use this result to obtain an analytical formula for the Hilbert-Schmidt
coherence of $n$-qudit states. We instantiate this general result
by looking at one-qubit (Sec. \ref{sec:qubit}), one-qutrit (Sec.
\ref{sec:qutrit}), two-qubit (Sec. \ref{sec:2qubits}), and two copies
of one-qubit (Sec. \ref{sec:rho_rho}) states. After writing, in Sec.
\ref{sec:2qubits}, the total HSC of two-qubit states in terms of
its local and nonlocal parts, in Sec. \ref{sec:rho_rho} we study
the possibility for controlling these quantities by manipulating the
local populations of two copies of one-qubit states. The contrasting
behaviors of HSC, $l_{1}$-norm coherence, and relative entropy of
coherence in this regard are noticed. We also investigate, in Sec.
\ref{sec:dynamics}, the dynamical behavior of quantum coherence functions
under the action of qutrit phase damping and amplitude damping channels.
Finally, the non-monotonicity of the HSD under tensor products is
investigated in Sec. \ref{sec:NMuTP}, where we report the first example
of an important implication of this kind of property of a quantum
distance function. We summarize our findings and discuss related open
questions in Sec. \ref{sec:conclusion}.

\section{Hilbert-Schmidt distance between $n$-qudit states}

\label{sec:hsd_nqudits}

Let $\mathbb{I}_{d_{s}}=\Gamma_{0}=\Gamma_{0}^{d}$ denote the $d_{s}\mathrm{x}d_{s}$
identity matrix, and $\{\Gamma_{j_{s}}\}_{j_{s}=1}^{d_{s}^{2}-1}$
be a basis for $SU(d_{s})$. We can use these operators to form a
basis for $SU(d)=\bigotimes_{s=1}^{n}SU(d_{s})$, with $d=\Pi_{s=1}^{n}d_{s}$,
and to write any $n$-qudit state as \cite{Kimura,Krammer-1}:
\begin{equation}
\rho=\sum_{j_{1}=0}^{d_{1}^{2}-1}\sum_{j_{2}=0}^{d_{2}^{2}-1}\cdots\sum_{j_{n}=0}^{d_{n}^{2}-1}r_{j_{1}j_{2}\cdots j_{n}}\Gamma_{j_{1}}\otimes\Gamma_{j_{2}}\otimes\cdots\otimes\Gamma_{j_{n}}.
\end{equation}
If the matrices $\{\Gamma_{j_{s}}\}_{j_{s}=0}^{d_{s}^{2}-1}$ satisfy
the algebraic relations
\begin{equation}
\mathrm{Tr}(\Gamma_{j_{s}}\Gamma_{k_{s}})=d_{s}^{\delta_{0j_{s}}}2^{1-\delta_{0j_{s}}}\delta_{j_{s}k_{s}},
\end{equation}
then the unit trace of $\rho$ requires $r_{00\cdots0}=1/d$. Besides,
the other parameters $r_{j_{1}j_{2}\cdots j_{n}}$, which are the
components of the so called Bloch's vector, can be obtained as follows:
\begin{eqnarray}
\mathrm{Tr}(\rho\Gamma_{k_{1}}\otimes\Gamma_{k_{2}}\otimes\cdots\otimes\Gamma_{k_{n}}) & = & \sum_{j_{1}=0}^{d_{1}^{2}-1}\sum_{j_{2}=0}^{d_{2}^{2}-1}\cdots\sum_{j_{n}=0}^{d_{n}^{2}-1}r_{j_{1}j_{2}\cdots j_{n}}\mathrm{Tr}(\Gamma_{j_{1}}\Gamma_{k_{1}})\mathrm{Tr}(\Gamma_{j_{2}}\Gamma_{k_{2}})\cdots\mathrm{Tr}(\Gamma_{j_{n}}\Gamma_{k_{n}})\nonumber \\
 & = & \sum_{j_{1}=0}^{d_{1}^{2}-1}\sum_{j_{2}=0}^{d_{2}^{2}-1}\cdots\sum_{j_{n}=0}^{d_{n}^{2}-1}r_{j_{1}j_{2}\cdots j_{n}}d_{1}^{\delta_{0j_{1}}}2^{1-\delta_{0j_{1}}}\delta_{j_{1}k_{1}}d_{2}^{\delta_{0j_{2}}}2^{1-\delta_{0j_{2}}}\delta_{j_{2}k_{2}}\cdots d_{n}^{\delta_{0j_{n}}}2^{1-\delta_{0j_{n}}}\delta_{j_{n}k_{n}}\nonumber \\
 & = & r_{k_{1}k_{2}\cdots k_{n}}2^{n-\sum_{s=1}^{n}\delta_{0k_{s}}}\prod_{s=1}^{n}d_{s}^{\delta_{0k_{s}}}.
\end{eqnarray}

Now let's regard the Hilbert-Schmidt distance (HSD) \cite{Cohen},
\begin{equation}
d_{hs}(\rho,\zeta)=\sqrt{\mathrm{Tr}(\rho-\zeta)^{2}},
\end{equation}
between two generic $n$-qudit states. Using $\rho-\zeta=\sum_{j_{1}=0}^{d_{1}^{2}-1}\cdots\sum_{j_{n}=0}^{d_{n}^{2}-1}(r_{j_{1}\cdots j_{n}}-z_{j_{1}\cdots j_{n}})\Gamma_{j_{1}}\otimes\cdots\otimes\Gamma_{j_{n}}$,
we shall have
\begin{eqnarray}
\mathrm{Tr}(\rho-\zeta)^{2} & = & \sum_{j_{1},k_{1}=0}^{d_{1}^{2}-1}\cdots\sum_{j_{n},k_{n}=0}^{d_{n}^{2}-1}(r_{j_{1}\cdots j_{n}}-z_{j_{1}\cdots j_{n}})(r_{k_{1}\cdots k_{n}}-z_{k_{1}\cdots k_{n}})\mathrm{Tr}(\Gamma_{j_{1}}\Gamma_{k_{1}})\cdots\mathrm{Tr}(\Gamma_{j_{n}}\Gamma_{k_{n}})\nonumber \\
 & = & \sum_{j_{1},k_{1}=0}^{d_{1}^{2}-1}\cdots\sum_{j_{n},k_{n}=0}^{d_{n}^{2}-1}(r_{j_{1}\cdots j_{n}}-z_{j_{1}\cdots j_{n}})(r_{k_{1}\cdots k_{n}}-z_{k_{1}\cdots k_{n}})d_{1}^{\delta_{0j_{1}}}2^{1-\delta_{0j_{1}}}\delta_{j_{1}k_{1}}\cdots d_{n}^{\delta_{0j_{n}}}2^{1-\delta_{0j_{n}}}\delta_{j_{n}k_{n}}\nonumber \\
 & = & \sum_{j_{1}=0}^{d_{1}^{2}-1}\sum_{j_{2}=0}^{d_{2}^{2}-1}\cdots\sum_{j_{n}=0}^{d_{n}^{2}-1}(r_{j_{1}j_{2}\cdots j_{n}}-z_{j_{1}j_{2}\cdots j_{n}})^{2}2^{n-\sum_{s=1}^{n}\delta_{0j_{s}}}\prod_{s=1}^{n}d_{s}^{\delta_{0j_{s}}}.
\end{eqnarray}
Thus, if we define the components of the \emph{rescaled Bloch's vector}
as
\begin{eqnarray}
R_{j_{1}j_{2}\cdots j_{n}} & =r_{j_{1}j_{2}\cdots j_{n}} & \sqrt{2^{n-\sum_{s=1}^{n}\delta_{0j_{s}}}{\textstyle \prod_{s=1}^{n}}d_{s}^{\delta_{0j_{s}}}}=\left(2^{n-\sum_{s=1}^{n}\delta_{0j_{s}}}\prod_{s=1}^{n}d_{s}^{\delta_{0j_{s}}}\right)^{-1/2}\mathrm{Tr}(\rho\Gamma_{j_{1}}\otimes\Gamma_{j_{2}}\otimes\cdots\otimes\Gamma_{j_{n}}),\label{eq:NBV}
\end{eqnarray}
with the analogous holding for $Z_{j_{1}j_{2}\cdots j_{n}}$, the
HSD between $\rho$ and $\zeta$ can be written as the Euclidean distance
between the $\mathbb{R}^{\Pi_{s=1}^{n}d_{s}^{2}-1}$ vectors $\vec{R}=\{R_{j_{1}j_{2}\cdots j_{n}}\}$
and $\vec{Z}=\{Z_{j_{1}j_{2}\cdots j_{n}}\}$:
\begin{equation}
d_{hs}(\rho,\zeta)=\sqrt{\sum_{j_{1}=0}^{d_{1}^{2}-1}\sum_{j_{2}=0}^{d_{2}^{2}-1}\cdots\sum_{j_{n}=0}^{d_{n}^{2}-1}(R_{j_{1}j_{2}\cdots j_{n}}-Z_{j_{1}j_{2}\cdots j_{n}})^{2}}=||\vec{R}-\vec{Z}||.\label{eq:HSD_nqudits}
\end{equation}

We observe that if the generators $\Gamma_{j_{s}}$ are Hermitian,
then the distinguishability of the two states gains a nice physical
significance in terms of differences between the corresponding pairs
of observables mean values. Notice that for this kind of generator
we have, e.g., $R_{j_{1}j_{2}\cdots j_{n}}=\left(2^{n-\sum_{s=1}^{n}\delta_{0j_{s}}}\prod_{s=1}^{n}d_{s}^{\delta_{0j_{s}}}\right)^{-1/2}\langle\Gamma_{j_{1}}\otimes\Gamma_{j_{2}}\otimes\cdots\otimes\Gamma_{j_{n}}\rangle_{\rho}$,
with $\langle X\rangle_{\rho}=\mathrm{Tr}(\rho X)$. For a generic
two-qudit density matrix, Ref. \cite{Maziero_AMP} showed an optimized
way to compute numerically the components of the associated Bloch's
vector (the Fortran code produced for that purpose can be accessed
freely in \cite{LibForQ}).

\section{Hilbert-Schmidt quantum coherence for $n$-qudit states}

\label{sec:hsc_nqudits}

Incoherent states, $\iota$, are represented by density matrices which
are diagonal in some reference, orthonormal, basis $\{|j\rangle\}_{j=1}^{d}$.
That is to say \cite{Plenio_QC}
\begin{equation}
\iota=\sum_{j=1}^{d}\iota_{j}|j\rangle\langle j|,
\end{equation}
with $\{\iota_{j}\}_{j=1}^{d}$ being a probability distribution.
The Hilbert-Schmidt coherence (HSC) of a state $\rho$ is defined
as its minimum Hilbert-Schmidt distance (HSD) to incoherent states:
\begin{equation}
C_{hs}(\rho)=\min_{\iota}d_{hs}(\rho,\iota).\label{eq:HSC}
\end{equation}
The HSC is nonnegative and is equal to zero if and only if $\rho$
is an incoherent state. It is also worthwhile mentioning that although
the HSC is not a ``full coherence monotone'' \cite{Plenio_QC}, it
is a monotone (i.e. $C_{hs}(\rho)\ge C_{hs}(\Delta(\rho))$) under
incoherent operations of the dephasing type (in basis $\{|j\rangle\}_{j=1}^{d}$):
$\Delta(\rho)=\sum_{j=1}^{d}|j\rangle\langle j|\rho|j\rangle\langle j|$
\cite{Cohen}.

For computing the HSC, one convenient choice for the $SU(d_{s})$
generators is the $d_{s}\mathrm{x}d_{s}$ generalized Gell Mann matrices
\cite{Krammer-1}:
\begin{eqnarray}
 &  & \Gamma_{j_{s}}^{d}=\sqrt{\frac{2}{j_{s}(j_{s}+1)}}\sum_{k_{s}=1}^{j_{s}+1}(-j_{s})^{\delta_{k_{s},j_{s}+1}}|k_{s}\rangle\langle k_{s}|\mbox{, }\mbox{ for }j_{s}=1,\cdots,d_{s}-1,\label{eq:SU1}\\
 &  & \Gamma_{(k_{s},l_{s})}^{s}=|k_{s}\rangle\langle l_{s}|+|l_{s}\rangle\langle k_{s}|,\mbox{ for }1\leq k_{s}<l_{s}\leq d_{s},\label{eq:SU2}\\
 &  & \Gamma_{(k_{s},l_{s})}^{a}=-i(|k_{s}\rangle\langle l_{s}|-|l_{s}\rangle\langle k_{s}|),\mbox{ for }1\leq k_{s}<l_{s}\leq d_{s}.\label{eq:SU3}
\end{eqnarray}
These three sets of matrices are named the \emph{diagonal}, \emph{symmetric},
and \emph{antisymmetric} sets, respectively. The particular cases
with $d_{s}=2$ and $d_{s}=3$ give the well known Pauli and Gell
Mann matrices, respectively. 

If the reference basis is chosen to be the $n$-qubit \emph{computational
basis} $\bigotimes_{s=1}^{n}|j_{s}\rangle$, with $j_{s}=1,\cdots,d_{s}$,
then the presence of any non-null component of the Bloch's vector
corresponding to generators from the symmetric and/or antisymmetric
sets implies in the presence of off diagonal elements in the density
matrix. Thus, the more general $n$-qudit incoherent state is obtained
utilizing the identity matrix and the generators from the diagonal
set:
\begin{equation}
\iota=\sum_{j_{1}=0}^{d_{1}-1}\sum_{j_{2}=0}^{d_{2}-1}\cdots\sum_{j_{n}=0}^{d_{n}-1}\iota_{j_{1}j_{2}\cdots j_{n}}\Gamma_{j_{1}}^{d}\otimes\Gamma_{j_{2}}^{d}\otimes\cdots\otimes\Gamma_{j_{n}}^{d}.
\end{equation}

As seen in Eq. (\ref{eq:HSD_nqudits}), the HSD is given in terms
of the sum of positive terms, which are equal to the square of the
difference between corresponding components of the rescaled vectors.
Therefore, the minimum in Eq. (\ref{eq:HSC}) is obtained if we set
$\iota_{j_{1}j_{2}\cdots j_{n}}=r_{j_{1}j_{2}\cdots j_{n}}$ for all
$j_{s}=0,\cdots,d_{s}-1$ and $s=1,2,\cdots,n$. Thus we see that
the optimal incoherent state under the HSD is
\begin{equation}
\iota_{\rho}=\sum_{j_{1}=0}^{d_{1}-1}\sum_{j_{2}=0}^{d_{2}-1}\cdots\sum_{j_{n}=0}^{d_{n}-1}r_{j_{1}j_{2}\cdots j_{n}}\Gamma_{j_{1}}^{d}\otimes\Gamma_{j_{2}}^{d}\otimes\cdots\otimes\Gamma_{j_{n}}^{d}.
\end{equation}

We shall define the \emph{coherence vector}, $\vec{C}$, as the vector
obtained using the components of the rescaled Bloch's vector in Eq.
(\ref{eq:NBV}) corresponding to one or more generators from the symmetric
and/or antisymmetric sets. Hence the HSC of a $n$-qudit state can
be cast as the Euclidean norm of this dimension $d(d-1)$ coherence
vector:
\begin{equation}
C_{hs}(\rho)=||\vec{C}||.
\end{equation}
In the next sub-sections, we exemplify the application of this general
result to some particular cases.

\subsection{One-qubit state}

\label{sec:qubit}

In this example $n=1$, $d_{1}=2$, and $\vec{C}=\left(\langle\Gamma_{(1,2)}^{s}\rangle,\langle\Gamma_{(1,2)}^{a}\rangle\right)/\sqrt{2}$.
So, the one-qubit HSC is given by
\begin{equation}
C_{hs}(\rho_{qb})=2^{-1/2}\sqrt{\langle\Gamma_{(1,2)}^{s}\rangle^{2}+\langle\Gamma_{(1,2)}^{a}\rangle^{2}}.\label{eq:C1qb}
\end{equation}
In this section, and in the remainder of this article, when there
is no chance for confusion, we omit the density matrix in the expression
for the mean values.

\subsection{One-qutrit state}

\label{sec:qutrit}

Here $n=1$, $d_{1}=3$, and $\vec{C}=\left(\langle\Gamma_{(1,2)}^{s}\rangle,\langle\Gamma_{(1,3)}^{s}\rangle,\langle\Gamma_{(2,3)}^{s}\rangle,\langle\Gamma_{(1,2)}^{a}\rangle,\langle\Gamma_{(1,3)}^{a}\rangle,\langle\Gamma_{(2,3)}^{a}\rangle\right)/\sqrt{2}$.
So,
\begin{eqnarray}
C_{hs}(\rho_{qt}) & = & 2^{-1/2}\sqrt{\langle\Gamma_{(1,2)}^{s}\rangle^{2}+\langle\Gamma_{(1,3)}^{s}\rangle^{2}+\langle\Gamma_{(2,3)}^{s}\rangle^{2}+\langle\Gamma_{(1,2)}^{a}\rangle^{2}+\langle\Gamma_{(1,3)}^{a}\rangle^{2}+\langle\Gamma_{(2,3)}^{a}\rangle^{2}}.
\end{eqnarray}

\subsection{Two-qubit state}

\label{sec:2qubits}

Now $n=2$, $d_{1}=2$, $d_{2}=2$ and
\begin{eqnarray}
C_{hs}^{2}(\rho_{2qb}) & = & 2^{-1}\sum_{j=0,1}\sum_{\kappa=s,a}\left(\langle\Gamma_{j}^{d}\otimes\Gamma_{(1,2)}^{\kappa}\rangle^{2}+\langle\Gamma_{(1,2)}^{\kappa}\otimes\Gamma_{j}^{d}\rangle^{2}\right)+2^{-1}\sum_{\kappa,\kappa'=s,a}\langle\Gamma_{(1,2)}^{\kappa}\otimes\Gamma_{(1,2)}^{\kappa'}\rangle^{2}\nonumber \\
 & = & C_{l}^{hs}(\rho_{2qb})+C_{nl}^{hs}(\rho_{2qb}),
\end{eqnarray}
where the local part of the coherence was set to
\begin{equation}
C_{l}^{hs}(\rho_{2qb})=2^{-1}\sum_{\kappa=s,a}\left(\langle\Gamma_{0}^{d}\otimes\Gamma_{(1,2)}^{\kappa}\rangle^{2}+\langle\Gamma_{(1,2)}^{\kappa}\otimes\Gamma_{0}^{d}\rangle^{2}\right)=C_{hs}^{2}(\rho_{2qb}^{(2)})+C_{hs}^{2}(\rho_{2qb}^{(1)}),
\end{equation}
with $\rho_{2qb}^{(s)}=\mathrm{Tr}_{\bar{s}}(\rho_{2qb})$ being the
one-qubit reduced states (their HSC read as in Eq. (\ref{eq:C1qb})).
The non-local part of the coherence \cite{Maziero_NLQC} may be cast
as
\begin{equation}
C_{nl}^{hs}(\rho_{2qb})=C_{hs}^{2}(\rho_{2qb})-C_{l}^{hs}(\rho_{2qb}).
\end{equation}

\subsection{Two copies of one-qubit states: Controlling coherences via the manipulation
of populations}

\label{sec:rho_rho}

As a special case of the class of states regarded in the last sub-section,
let us look to two copies of one-qubit states: $\rho_{qb}\otimes\rho_{qb}$.
For this kind of state we have $C_{l}^{hs}(\rho_{qb}\otimes\rho_{qb})=2C_{hs}^{2}(\rho_{qb})$
and
\begin{eqnarray}
C_{nl}^{hs}(\rho_{qb}\otimes\rho_{qb}) & = & 2^{-1}\sum_{\kappa=s,a}\left(\langle\Gamma_{1}^{d}\otimes\Gamma_{(1,2)}^{\kappa}\rangle_{\rho_{qb}\otimes\rho_{qb}}^{2}+\langle\Gamma_{(1,2)}^{\kappa}\otimes\Gamma_{1}^{d}\rangle_{\rho_{qb}\otimes\rho_{qb}}^{2}\right)+2^{-1}\sum_{\kappa,\kappa'=s,a}\langle\Gamma_{(1,2)}^{\kappa}\otimes\Gamma_{(1,2)}^{\kappa'}\rangle_{\rho_{qb}\otimes\rho_{qb}}^{2}\nonumber \\
 & = & \langle\Gamma_{1}^{d}\rangle^{2}\sum_{\kappa=s,a}\langle\Gamma_{(1,2)}^{\kappa}\rangle_{\rho_{qb}}^{2}+2^{-1}\sum_{\kappa,\kappa'=s,a}\langle\Gamma_{(1,2)}^{\kappa}\rangle_{\rho_{qb}}^{2}\langle\Gamma_{(1,2)}^{\kappa'}\rangle_{\rho_{qb}}^{2}\nonumber \\
 & = & 2C_{hs}^{2}(\rho_{qb})\left(\langle\Gamma_{1}^{d}\rangle_{\rho_{qb}}^{2}+C_{hs}^{2}(\rho_{qb})\right).
\end{eqnarray}

Considering that the local populations are equal to
\begin{equation}
p_{j}=\mathrm{Tr}(\rho_{qb}|j\rangle\langle j|)=(1+(-1)^{\delta_{j2}}\langle\Gamma_{1}^{d}\rangle)/2,
\end{equation}
with $j=1,2$, we notice that although we cannot control the local
HSC by manipulating $p_{j}$, we can change the non-local HSC by modifying
the difference between those populations: $\langle\Gamma_{1}^{d}\rangle=p_{1}-p_{2}$.

The result above is to be compared with the one for the $l_{1}$-norm
coherence (L1C), which, in contrast to the HSC, is a coherence monotone
under the incoherent operations regarded in Ref. \cite{Plenio_QC}.
The L1C is given by the sum of the absolute values of the off-diagonal
elements of a density matrix. Although $\langle\Gamma_{1}^{d}\rangle$
is present in the off-diagonal elements of $\rho_{qb}\otimes\rho_{qb}$,
it cancels out when we perform that sum, and we get 
\begin{equation}
C_{l_{1}}(\rho_{qb}\otimes\rho_{qb})=\left(1+C_{l_{1}}(\rho_{qb})\right)^{2}-1,
\end{equation}
with $C_{l_{1}}(\rho_{qb})=2^{1/2}C_{hs}(\rho_{qb})$. So, we cannot
control the non-local L1C by tuning local populations. 

\begin{figure}
\begin{centering}
\includegraphics[scale=0.4]{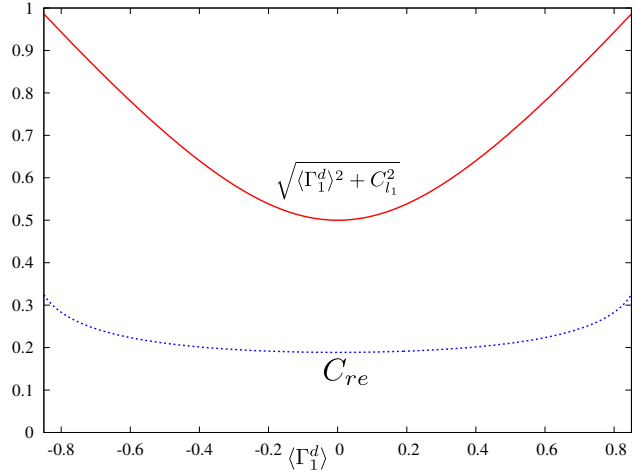}
\par\end{centering}

\caption{(color online) Relative entropy of coherence as a function of the
qubit's population-controlling parameter $\langle\Gamma_{1}^{d}\rangle$
for a fixed value of the $l_{1}$-norm coherence $C_{l_{1}}(\rho_{qb})=\sqrt{\langle\Gamma_{(1,2)}^{s}\rangle^{2}+\langle\Gamma_{(1,2)}^{a}\rangle^{2}}=\sqrt{0+1/4}.$
If we look at the Bloch's sphere in the $\mathbb{R}^{3}$ space $(\langle\Gamma_{1}^{d}\rangle,\langle\Gamma_{(1,2)}^{s}\rangle,\langle\Gamma_{(1,2)}^{a}\rangle)$,
a constant value of $C_{l_{1}}$ (or of $C_{hs}$) will be associated
with points in the lateral surface of a cylinder of physical states
within the sphere. In this cylinder, for $\langle\Gamma_{1}^{d}\rangle$
running from $-\sqrt{1-C_{l_{1}}^{2}}$ to $\sqrt{1-C_{l_{1}}^{2}}$,
different values of $|\langle\Gamma_{1}^{d}\rangle|$ shall lead to
different values of $C_{re}$. The uppermost and downmost circles
of that cylinder surface touch the sphere surface, and in these circles
lie pure states for which the Bloch's vector norm, shown in the upper
(red) curve, is equal to one.}

\label{fig:REC}
\end{figure}

Because of its operational interpretation \cite{Yang_RTC}, the relative
entropy of coherence (REC) may be considered to be one of the main
coherence quantifiers introduced up today. The REC is defined and
given by \cite{Plenio_QC}:
\begin{equation}
C_{re}(\rho)=\min_{\iota}d_{re}(\rho,\iota)=\min_{\iota}(-\mathrm{Tr}(\rho\log_{2}\iota)-S(\rho))=S(\rho_{diag})-S(\rho),
\end{equation}
with $\rho_{diag}$ being the density matrix obtained by setting the
off-diagonal elements of $\rho$ to zero and $S(x)=-\mathrm{Tr}(x\log_{2}x)$
is von Neumann's entropy. Using the additivity of von Neumann's entropy
under tensor products, i.e., $S(x\otimes x)=2S(x)$ and the fact that
$(\rho\otimes\rho)_{diag}=\rho_{diag}\otimes\rho_{diag}$, we see
that the REC is also additive under tensor products, i.e., $C_{re}(\rho\otimes\rho)=2C_{re}(\rho)$.
So, we shall be able to change the REC of the two copies of $\rho$
only if we can change its own REC. An interesting result we report
here in this direction is that, contrary to what happens with the
HSC and L1C, we can modify the REC of a qubit by controlling its populations,
a quantity usually considered to have a classical character. For a
qubit state, $\rho_{qb}=2^{-1}(\Gamma_{0}+\langle\Gamma_{1}^{d}\rangle\Gamma_{1}^{d}+\langle\Gamma_{(1,2)}^{s}\rangle\Gamma_{(1,2)}^{s}+\langle\Gamma_{(1,2)}^{a}\rangle\Gamma_{(1,2)}^{a})$,
\begin{equation}
C_{re}(\rho_{qb})=-\sum_{j=1}^{2}\frac{1+(-1)^{\delta_{j2}}\langle\Gamma_{1}^{d}\rangle}{2}\log_{2}\frac{1+(-1)^{\delta_{j2}}\langle\Gamma_{1}^{d}\rangle}{2}+\sum_{j=1}^{2}\frac{1+(-1)^{\delta_{j2}}G}{2}\log_{2}\frac{1+(-1)^{\delta_{j2}}G}{2},
\end{equation}
with $G^{2}=$$\langle\Gamma_{1}^{d}\rangle^{2}+C_{l_{1}}^{2}(\rho_{qb}).$
The form of the equation for $C_{re}(\rho_{qb})$ is an indicative
that the term $\langle\Gamma_{1}^{d}\rangle$, related to the populations,
does not cancel out. We verify that this is indeed the case via an
example, which is shown in Fig. \ref{fig:REC}.

\section{Coherence dynamics under qutrit dephasing and dissipation}

\label{sec:dynamics}

In this section, we study the dynamics of some coherence functions
under decoherence. Once for \emph{one-qubit} states the Hilbert-Schmidt
coherence (HSC) and the $l_{1}$-norm coherence (L1C) are proportional,
their dynamic behavior shall be equivalent in this physically meaningful
kind of process. For instance, for a qubit the freezing conditions
\cite{Maziero_NLQC,Adesso_frozenC} for $C_{hs}$ and for $C_{l_{1}}$
are the same. However, even in this simplest context, the dynamic
behavior of relative entropy of coherence (REC) is qualitatively disparate
from those of HSC and L1C. And this is due to the dependence of the
first on the populations. On the other hand, already for \emph{qutrits}
the proportionality between $C_{hs}$ and $C_{l_{1}}$ does not hold
in general. Here we have $C_{l_{1}}(\rho_{qt})=\sum_{k,l}\sqrt{\langle\Gamma_{(k,l)}^{s}\rangle^{2}+\langle\Gamma_{(k,l)}^{a}\rangle^{2}}$
with $1\le k<l\le3$. So, we can warrant that $C_{l_{1}}\propto C_{hs}$
only if one single pair $(\langle\Gamma_{(k,l)}^{s}\rangle,\langle\Gamma_{(k,l)}^{a}\rangle)$
is non-null. In the following we shall regard the dynamics induced
by qutrit phase damping (PD) and amplitude damping (AD) channels.
The Kraus' operators for these channels are (see e.g. Refs. \cite{Guo_qutrit,Khan_qutrit}
and references therein):
\begin{eqnarray}
 &  & K_{0}^{pd}=\sqrt{1-p}\mathbb{I}_{3}\mbox{ and }K_{j}^{pd}=\sqrt{p}|j\rangle\langle j|\mbox{ for }j=1,2,3,\\
 &  & K_{0}^{ad}=|1\rangle\langle1|+\sqrt{1-p}(|2\rangle\langle2|+|3\rangle\langle3|)\mbox{, }K_{1}^{ad}=\sqrt{p}|1\rangle\langle2|\mbox{, }K_{2}^{ad}=\sqrt{p}|1\rangle\langle3|,
\end{eqnarray}
with $\mathbb{I}_{3}$ being the $3\mathrm{x}3$ identity matrix. 

\begin{figure}
\includegraphics[scale=0.42]{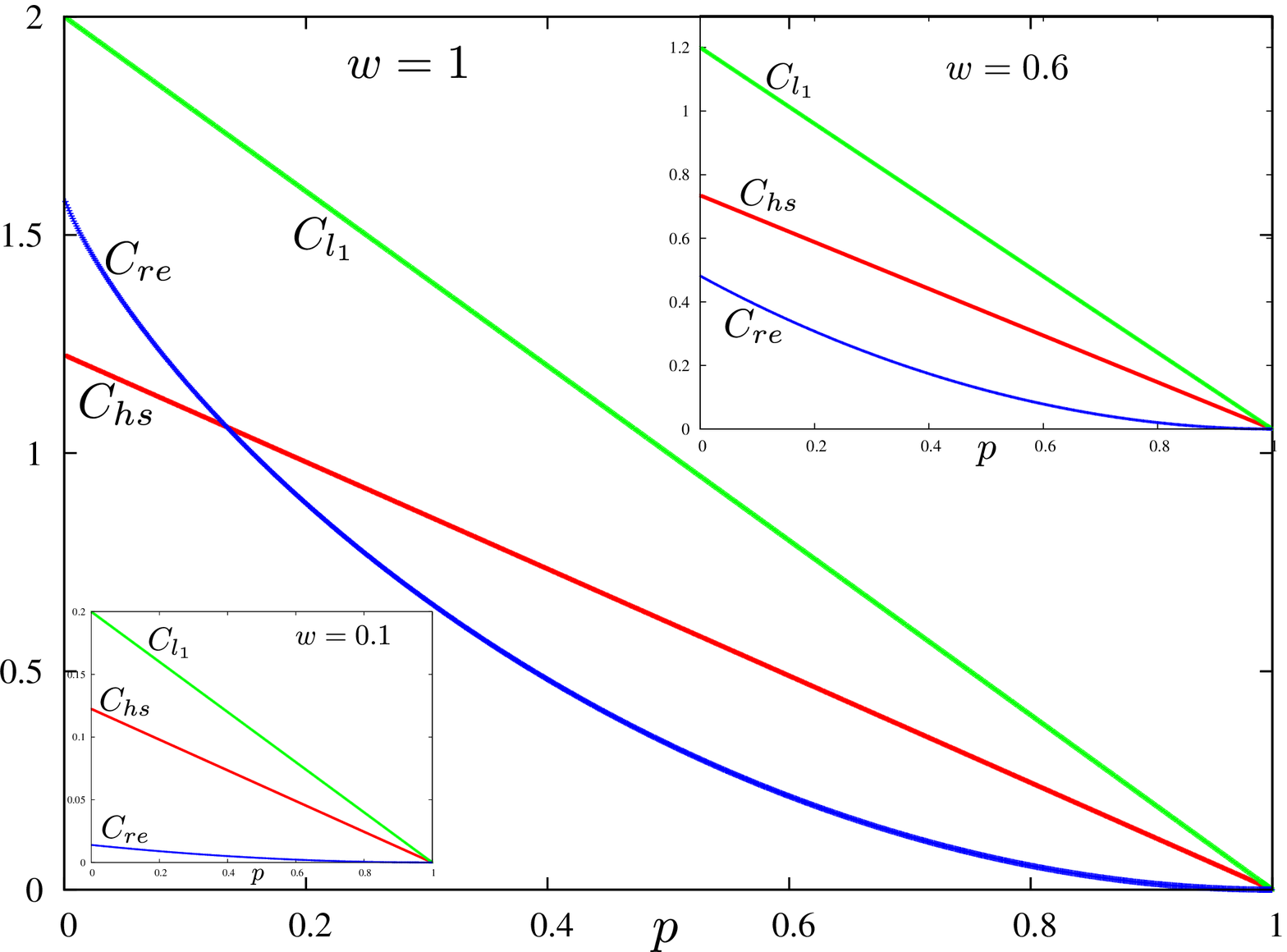}\hspace{0.25cm}\includegraphics[scale=0.42]{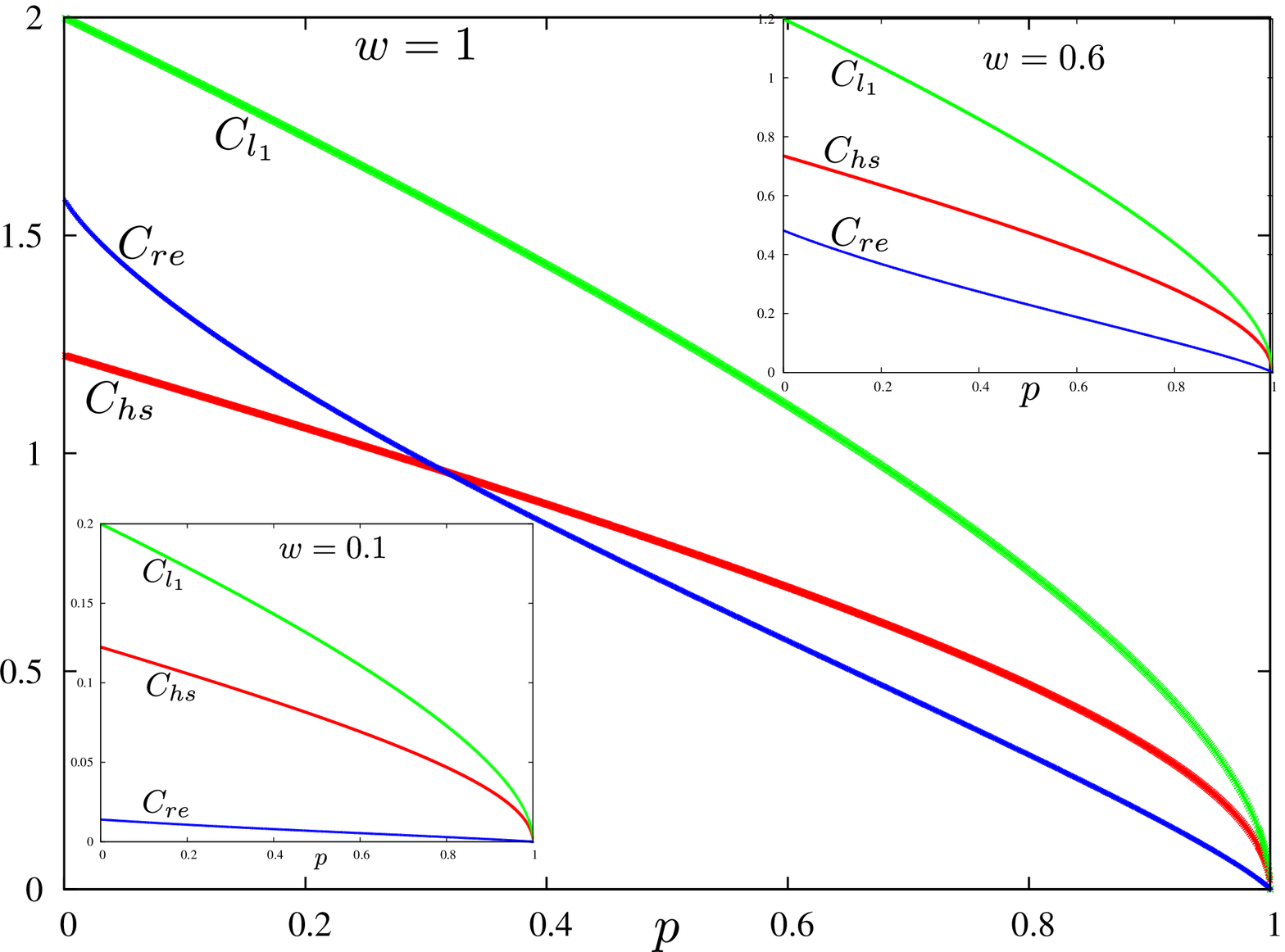}

\caption{(color online) Coherence of Hilbert-Schmidt (HSC), of the $l_{1}$-norm
(L1C), and of the relative entropy (REC) under the action of the phase
damping channel (plots on the left) and of the amplitude damping channel
(plots on the right) for initial states $\rho_{w}=(1-w)\mathbb{I}_{3}/3+w|\psi\rangle\langle\psi|$
with $|\psi\rangle=3^{-1/2}\sum_{j=1}^{3}|j\rangle$. For these initial
states, the dynamical behavior of HSC and L1C is seen to be similar,
but quite different from that of the REC (regarding the functions
concavities).}

\label{fig:qtdyn}
\end{figure}

For a qutrit prepared in a general state and submitted to the action
of the PD or AD channels, the states at parametrized time $p$ \cite{Soares-Pinto2011},
$\rho_{p}=\sum_{j}K_{j}\rho K_{j}^{\dagger}$, are given as follows:
\begin{eqnarray}
 &  & \rho_{p}^{pd}=\begin{bmatrix}3^{-1}+(\langle\Gamma_{1}^{d}\rangle+\langle\Gamma_{2}^{d}\rangle)/2 & q(\langle\Gamma_{(1,2)}^{s}\rangle-i\langle\Gamma_{(1,2)}^{a}\rangle)/2 & q(\langle\Gamma_{(1,3)}^{s}\rangle-i\langle\Gamma_{(1,3)}^{a}\rangle)/2\\
q(\langle\Gamma_{(1,2)}^{s}\rangle+i\langle\Gamma_{(1,2)}^{a}\rangle)/2 & 3^{-1}-(\langle\Gamma_{1}^{d}\rangle-\langle\Gamma_{2}^{d}\rangle)/2 & q(\langle\Gamma_{(2,3)}^{s}\rangle-i\langle\Gamma_{(2,3)}^{a}\rangle)/2\\
q(\langle\Gamma_{(1,3)}^{s}\rangle+i\langle\Gamma_{(1,3)}^{a}\rangle)/2 & q(\langle\Gamma_{(2,3)}^{s}\rangle+i\langle\Gamma_{(2,3)}^{a}\rangle)/2 & 3^{-1}-\langle\Gamma_{2}^{d}\rangle
\end{bmatrix},\\
 &  & \rho_{p}^{ad}=\begin{bmatrix}(1+2p)/3+q(\langle\Gamma_{1}^{d}\rangle+\langle\Gamma_{2}^{d}\rangle)/2 & \sqrt{q}(\langle\Gamma_{(1,2)}^{s}\rangle-i\langle\Gamma_{(1,2)}^{a}\rangle)/2 & \sqrt{q}(\langle\Gamma_{(1,3)}^{s}\rangle-i\langle\Gamma_{(1,3)}^{a}\rangle)/2\\
\sqrt{q}(\langle\Gamma_{(1,2)}^{s}\rangle+i\langle\Gamma_{(1,2)}^{a}\rangle)/2 & q(3^{-1}-(\langle\Gamma_{1}^{d}\rangle-\langle\Gamma_{2}^{d}\rangle)/2) & q(\langle\Gamma_{(2,3)}^{s}\rangle-i\langle\Gamma_{(2,3)}^{a}\rangle)/2\\
\sqrt{q}(\langle\Gamma_{(1,3)}^{s}\rangle+i\langle\Gamma_{(1,3)}^{a}\rangle)/2 & q(\langle\Gamma_{(2,3)}^{s}\rangle+i\langle\Gamma_{(2,3)}^{a}\rangle)/2 & q(3^{-1}-\langle\Gamma_{2}^{d}\rangle)
\end{bmatrix},
\end{eqnarray}
with $q=1-p$. As all off-diagonal elements of $\rho_{p}^{pd}$ decay
at the same rate, we see that for the PD channel (PDC) the following
motion equation holds for the L1C: $C_{l_{1}}(\rho_{p}^{pd})=(1-p)C_{l_{1}}(\rho^{pd})$.
Besides, as all Bloch's vector components from the symmetric and anti-symmetric
sets are proportional to $1-p$, we get the following evolution equation
for HSC: $C_{hs}(\rho_{p}^{pd})=(1-p)C_{hs}(\rho^{pd})$. Therefore,
even if there is no simple-general proportionality relation between
the initial values of L1C and HSC, the rate with which these values
diminish with time for the PDC is once again observed to be equivalent.
On the other hand, for this particular model of dissipation of three
level systems, not all off-diagonal elements of $\rho_{p}^{ad}$ change
equally with time. Here the time dependence of the coherences for
the AD channel are:
\begin{eqnarray}
C_{l_{1}}(\rho_{p}^{ad}) & = & \sqrt{1-p}\left(\sqrt{\langle\Gamma_{(1,2)}^{s}\rangle^{2}+\langle\Gamma_{(1,2)}^{a}\rangle^{2}}+\sqrt{\langle\Gamma_{(1,3)}^{s}\rangle^{2}+\langle\Gamma_{(1,3)}^{a}\rangle^{2}}\right)+(1-p)\sqrt{\langle\Gamma_{(2,3)}^{s}\rangle^{2}+\langle\Gamma_{(2,3)}^{a}\rangle^{2}},\\
C_{hs}(\rho_{p}^{ad}) & = & \frac{1}{\sqrt{2}}\sqrt{(1-p)\left(\langle\Gamma_{(1,2)}^{s}\rangle^{2}+\langle\Gamma_{(1,2)}^{a}\rangle^{2}+\langle\Gamma_{(1,3)}^{s}\rangle^{2}+\langle\Gamma_{(1,3)}^{a}\rangle^{2}\right)+(1-p)^{2}\left(\langle\Gamma_{(2,3)}^{s}\rangle^{2}+\langle\Gamma_{(2,3)}^{a}\rangle^{2}\right)}.
\end{eqnarray}
We illustrate the time dependence of these functions, and of REC,
with the specific examples shown in Fig. \ref{fig:qtdyn}. 

\section{Non-monotonicity of the HSD under tensor products}

\label{sec:NMuTP}

\begin{figure}
\begin{centering}
\includegraphics[scale=0.5]{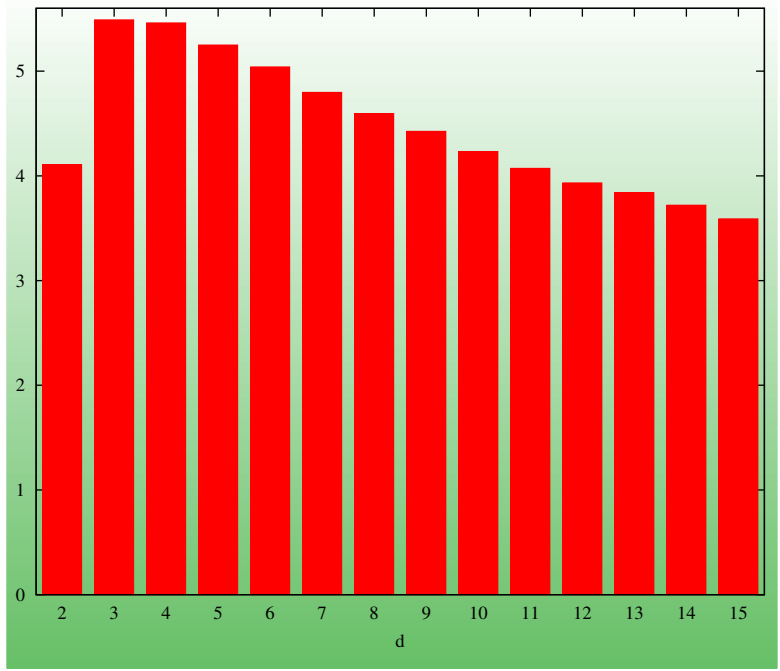}
\par\end{centering}

\caption{(color online) Percentage of the random quartets of states leading
to the non-monotonic behavior under tensor products of the Hilbert-Schmidt
distance. Using the standard method from the library described in
Ref. \cite{Maziero_LibForro} (the associated Fortran code can be
accessed freely in \cite{LibForro}), we generated one million quartets
of states for each dimension $d$.}

\label{fig:nmutp}
\end{figure}

One can verify that, similarly to what happens with the trace distance
(TD) \cite{Maziero_NMuTP}, the Hilbert-Schmidt distance (HSD) does
not suffer from the non-monotonicity under tensor product (NMuTP)
issue for general pure states or for one-qubit collinear states. Actually,
one can show that this holds true for all $p$-norm distances \cite{Ruskai}.
For general qudit states, the probability for random quartets of states
to present the NMuTP behavior of HSD decays with increasing the system
dimension, also similar to what was seen for the TD \cite{Maziero_NMuTP}.
For completeness, we show such probabilities in Fig. \ref{fig:nmutp}.

Regarding the Hilbert-Schmidt coherence (HSC), once
\begin{equation}
C_{hs}^{2}(\rho_{qb}\otimes\rho_{qb})=2C_{hs}^{2}(\rho_{qb})\left(1+C_{hs}^{2}(\rho_{qb})+\langle\Gamma_{1}^{d}\rangle_{\rho_{qb}}\right),
\end{equation}
we see that the NMuTP behavior of the HSD can be revealed on its associated
HSC if we manipulate $\langle\Gamma_{1}^{d}\rangle_{\rho_{qb}}$ accordingly;
what we will do by using an example. We will consider $\langle\Gamma_{(1,2)}^{s}\rangle_{\rho_{qb}}=\langle\Gamma_{(1,2)}^{a}\rangle_{\rho_{qb}}=0.34$
and $\langle\Gamma_{(1,2)}^{s}\rangle_{\xi_{qb}}=\langle\Gamma_{(1,2)}^{a}\rangle_{\xi_{qb}}=0.33$,
what gives us $C_{hs}(\rho_{qb})=0.34$ and $C_{hs}(\xi_{qb})=0.33,$
so
\begin{equation}
C_{hs}(\rho_{qb})>C_{hs}(\xi_{qb}).
\end{equation}
In order to invert the dissimilarity relation, we fix $\langle\Gamma_{1}^{d}\rangle_{\rho_{qb}}=0$
and change $|\langle\Gamma_{1}^{d}\rangle_{\xi_{qb}}|$, which can
vary from zero up to $\sqrt{1-2C_{hs}^{2}(\xi_{qb})}$. As $C_{hs}(\rho_{qb}\otimes\rho_{qb})\approx0.26$,
we can use e.g. $|\langle\Gamma_{1}^{d}\rangle_{\xi_{qb}}|=0.7$ to
get $C_{hs}(\xi_{qb}\otimes\xi_{qb})\approx0.39$, and consequently
\begin{equation}
C_{hs}(\rho_{qb}\otimes\rho_{qb})<C_{hs}(\xi_{qb}\otimes\xi_{qb}).
\end{equation}

These last two inequalities are a manifestation of the NMuTP of HSD:
\begin{equation}
d(\rho,\iota_{\rho})>d(\xi,\iota_{\xi})\mbox{ and }d(\rho\otimes\rho,\iota_{\rho}\otimes\iota_{\rho})<d(\xi\otimes\xi,\iota_{\xi}\otimes\iota_{\xi}).
\end{equation}
In the last equation we used the fact that, under the HSD, $\iota_{x\otimes x}=\iota_{x}\otimes\iota_{x}$.

\section{Concluding remarks}

\label{sec:conclusion}

In this article, we started obtaining a closed formula for the Hilbert-Schmidt
distance (HSD) between two generic $n$-qudit states in terms of the
Euclidean distance between the corresponding rescaled Bloch's vectors.
We used this result to derive an analytical expression for the Hilbert-Schmidt
quantum coherence (HSC) of $n$-qudit systems. This formula was then
exemplified in simple cases and we applied it to investigated issues
related to the distribution and transformation of quantum coherence
in some composite systems.

After splitting the total HSC of two-qubit states into its local and
non-local parts, we analyzed the possibility of controlling these
coherences by tuning the local populations of two copies of a generic
one-qubit state. Related to that matter, we observed some interesting
contrasting behaviors of the HSC, $l_{1}$-norm coherence (L1C), and
relative entropy of coherence (REC). We noticed that it is not possible
to change the local HSC and L1C by tuning local populations. However,
contrary the what was seen for the L1C, we can control the HSC by
doing that kind of operation. The REC of two copies of a qubit state
can be changed only if the local REC is changed, what we showed being
possible by the control of local populations. This raises yet another
bold difference among these coherence functions. In future works,
it could be fruitful to investigate this kind of issue considering
its possible physical implications and also another coherence quantifiers. 

We also investigated the time evolution of coherence under qutrit
dephasing and dissipation environments. The population-dependent character
of the REC leads to its quite different dynamical behavior when compared
to those of L1C and HSC. These last two coherence functions are seen
to be equivalent for qubits and for some qutrit states. Besides, we
have shown that even when HSC and L1C are not proportional, their
dynamical behavior can be similar.

We have identified the first important implication of the non-monotonicity
under tensor product (NMuTP) property of a quantum distance measure.
We showed that the NMuTP of HSD, when applied to some states, can
indicate that although a configuration $\rho_{qb}$ is more coherent
than another state $\xi_{qb}$, two independent copies of the first
state possess less HSC than two copies of the last. We emphasize that
we succeeded in doing that at this moment mainly because we used the
HSD. This unexpected result, obtained for the HSD, calls for further
investigations of the NMuTP issue regarding other quantum distance
measures and also considering other of its possible consequences.
\begin{acknowledgments}
This work was supported by the Brazilian funding agencies: Conselho Nacional de Desenvolvimento Cient\'ifico e Tecnol\'ogico (CNPq), processes 441875/2014-9 and 303496/2014-2, Coordena\c{c}\~ao de Desenvolvimento de Pessoal de N\'ivel Superior (CAPES), process 6531/2014-08, and Instituto Nacional de Ci\^encia e Tecnologia de Informa\c{c}\~ao Qu\^antica (INCT-IQ), process 2008/57856-6.\end{acknowledgments}


\begin{thebibliography}{10}
\bibitem{Feynman} R. P. Feynman, R. B. Leighton, and M. Sands, \emph{The
Feynman Lectures on Physics}, Volume 3 (Addison-Wesley, Massachusetts,
1965). 

\bibitem{Wolf} L. Mandel and E. Wolf, \emph{Optical Coherence and
Quantum Optics} (Cambridge University Press, Cambridge, England, 2008).

\bibitem{Adesso_CE} A. Streltsov, U. Singh, H. S. Dhar, M. N. Bera,
and G. Adesso, Measuring quantum coherence with entanglement, Phys.
Rev. Lett. 115, 020403 (2015). 

\bibitem{Gu} J. Ma, B. Yadin, D. Girolami, V. Vedral, and M. Gu,
Converting coherence to quantum correlations, Phys. Rev. Lett. 116,
160407 (2016). 

\bibitem{Allegra} P. Giorda and M. Allegra, Coherence in quantum
estimation, arXiv:1611.02519.

\bibitem{Scholes} A. Chenu and G. D. Scholes, Coherence in energy
transfer and photosynthesis, Annu. Rev. Phys. Chem. 66, 69 (2015). 

\bibitem{Uzdin} R. Uzdin, Coherence-induced reversibility and collective
operation of quantum heat machines via coherence recycling, Phys.
Rev. Applied 6, 024004 (2016). 

\bibitem{Hillery} M. Hillery, Coherence as a resource in decision
problems: The Deutsch-Jozsa algorithm and a variation, Phys. Rev.
A 93, 012111 (2016). 

\bibitem{Ma} X. Yuan, K. Liu, Y. Xu, W. Wang, Y. Ma, F. Zhang, Z.
Yan, R. Vijay, L. Sun, and X. Ma, Experimental quantum randomness
processing using superconducting qubits, Phys. Rev. Lett. 117, 010502
(2016). 

\bibitem{Lewenstein-1} A. Streltsov, E. Chitambar, S. Rana, M. N.
Bera, A. Winter, and M. Lewenstein, Entanglement and coherence in
quantum state merging, Phys. Rev. Lett. 116, 240405 (2016). 

\bibitem{Pati} A. Misra, U. Singh, S. Bhattacharya, and A. K. Pati,
Energy cost of creating quantum coherence, Phys. Rev. A 93, 052335
(2016). 

\bibitem{Fan} H.-L. Shi, S.-Y. Liu, X.-H. Wang, W.-L. Yang, Z.-Y.
Yang, and H. Fan, Coherence depletion in the Grover quantum search
algorithm, Phys. Rev. A 95, 032307 (2017). 

\bibitem{Plenio_QC_RMP} A. Streltsov, G. Adesso, and M. B. Plenio,
Quantum coherence as a resource, arXiv:1609.02439. 

\bibitem{Gour} E. Chitambar and G. Gour, Critical examination of
incoherent operations and a physically consistent resource theory
of quantum coherence, Phys. Rev. Lett. 117, 030401 (2016). 

\bibitem{Spekkens} I. Marvian and R. W. Spekkens, How to quantify
coherence: Distinguishing speakable and unspeakable notions, Phys.
Rev. A 94, 052324 (2016). 

\bibitem{Plenio_QC} T. Baumgratz, M. Cramer, and M. B. Plenio, Quantifying
coherence, Phys. Rev. Lett. 113, 170401 (2014). 

\bibitem{Adesso_RoC} C. Napoli, T. R. Bromley, M. Cianciaruso, M.
Piani, N. Johnston, and G. Adesso, Robustness of coherence: An operational
and observable measure of quantum coherence, Phys. Rev. Lett. 116,
150502 (2016). 

\bibitem{Adesso_RoC-1} M. Piani, M. Cianciaruso, T. R. Bromley, C.
Napoli, N. Johnston, and G. Adesso, Robustness of asymmetry and coherence
of quantum states, Phys. Rev. A 93, 042107 (2016). 

\bibitem{Mintert} F. Levi and F. Mintert, A quantitative theory of
coherent delocalization, New J. Phys. 16, 033007 (2014). 

\bibitem{Ruskai} D. P\'erez-Garc\'ia, M. M. Wolf, D. Petz, and M.
B. Ruskai, Contractivity of positive and trace-preserving maps under
Lp norms, J. Math. Phys. 47, 083506 (2006).

\bibitem{Schirmer} X. Wang and S. G. Schirmer, Contractivity of the
Hilbert-Schmidt distance under open-system dynamics, Phys. Rev. A
79, 052326 (2009).

\bibitem{Hanggi} J. Dajka, J. {\L}uczka, and P. H\"anggi, Distance
between quantum states in the presence of initial qubit-environment
correlations: A comparative study, Phys. Rev. A 84, 032120 (2011). 

\bibitem{Ozawa} M. Ozawa, Entanglement measures and the Hilbert-Schmidt
distance, Phys. Lett. A 268, 158 (2000).

\bibitem{Piani} M. Piani, The problem with the geometric discord,
Phys. Rev. A 86, 034101 (2012).

\bibitem{Walther} B. Dakic, Y. O. Lipp, X. Ma, M. Ringbauer, S. Kropatschek,
S. Barz, T. Paterek, V. Vedral, A. Zeilinger, C. Brukner, and P. Walther,
Quantum discord as resource for remote state preparation, Nat. Phys.
8, 666 (2012).

\bibitem{Zhou} X. Wu and T. Zhou, Geometric discord: A resource for
increments of quantum key generation through twirling, Sci. Rep. 5,
13365 (2015). 

\bibitem{Thirring} R. A. Bertlmann, H. Narnhofer, and W. Thirring,
A geometric picture of entanglement and Bell inequalities, Phys. Rev.
A 66, 032319 (2002).

\bibitem{Krammer} R. A. Bertlmann, K. Durstberger, B. C. Hiesmayr,
and P. Krammer, Optimal entanglement witnesses for qubits and qutrits,
Phys. Rev. A 72, 052331 (2005).

\bibitem{Brukner} J. Lee, M. S. Kim, and C. Brukner, Operationally
invariant measure of the distance between quantum states by complementary
measurements, Phys. Rev. Lett. 91, 087902 (2003).

\bibitem{Cohen} B. Tamir and E. Cohen, A Holevo-type bound for a
Hilbert Schmidt distance measure, J. Quant. Inf. Science 05, 127 (2015).

\bibitem{Wunsche} V. V. Dodonov, O. V. Man\textquoteright{}ko, V.
I. Man\textquoteright{}ko, and A. W\"unsche, Hilbert-Schmidt distance
and non-classicality of states in quantum optics, J. Mod. Opt. 47,
633 (2000).

\bibitem{Sommers} K. Zyczkowski and H.-J. Sommers, Hilbert-Schmidt
volume of the set of mixed quantum states, J. Phys. A: Math. Gen.
36, 10115 (2003).

\bibitem{Winter} S. Popescu, A. J. Short, and A. Winter, Entanglement
and the foundations of statistical mechanics, Nat. Phys. 2, 754 (2006).

\bibitem{Soto} G. Bj\"ork, H. de Guise, A. B. Klimov, P. de la Hoz,
and L. L. S\'anchez-Soto, Classical distinguishability as an operational
measure of polarization, Phys. Rev. A 90, 013830 (2014). 

\bibitem{Trucks} C. Witte and M. Trucks, A new entanglement measure
induced by the Hilbert-Schmidt norm, Phys. Lett. A 257, 14 (1999). 

\bibitem{Moor} F. Verstraete, J. Dehaene, and B. De Moor, On the
geometry of entangled states, J. Mod. Opt. 49, 1277 (2002).

\bibitem{Maziero_HSE} J. Maziero, Computing partial transposes and
related entanglement functions, Braz. J. Phys. 46, 605 (2016). 

\bibitem{Brukner_HSDisc} B. Dakic, V. Vedral, and C. Brukner, Necessary
and sufficient condition for non-zero quantum discord, Phys. Rev.
Lett. 105, 190502 (2010). 

\bibitem{Li} L. Li, Q.-W. Wang, S.-Q. Shen, and M. Li, Geometric
measure of quantum discord with weak measurements, Quantum Inf. Process.
15, 291 (2016). 

\bibitem{Fu_HSDisc} S. Luo and S. Fu, Geometric measure of quantum
discord, Phys. Rev. A 82, 034302 (2010). 

\bibitem{Azmi} S. J. Akhtarshenas, H. Mohammadi, S. Karimi, and Z.
Azmi, Computable measure of quantum correlation, Quantum Inf. Process.
14, 247 (2015). 

\bibitem{Fu_MIN} S. Luo and S. Fu, Measurement-induced nonlocality,
Phys. Rev. Lett. 106, 120401 (2011). 

\bibitem{Adesso_NegHSD} D. Girolami and G. Adesso, Interplay between
computable measures of entanglement and other quantum correlations,
Phys. Rev. A 84, 052110 (2011). 

\bibitem{Hou} Y.-L. Yuan and X.-W. Hou, Thermal geometric discords
in a two-qutrit system, Int. J. Quantum Inform. 14, 1650016 (2016). 

\bibitem{Maziero_NLQC} M. B. Pozzobom and J. Maziero, Environment-induced
quantum coherence spreading of a qubit, Ann. Phys. 377, 243 (2017).

\bibitem{Byrnes} R. Chandrashekar, P. Manikandan, J. Segar, and T.
Byrnes, Distribution of quantum coherence in multipartite systems,
Phys. Rev. Lett. 116, 150504 (2016). 

\bibitem{Jeong} K. C. Tan, H. Kwon, C.-Y. Park, and H. Jeong, Unified
view of quantum correlations and quantum coherence, Phys. Rev. A 94,
022329 (2016).

\bibitem{Verstraete} K. M. R. Audenaert, J. Calsamiglia, R. Mun\~oz-Tapia,
E. Bagan, Ll. Masanes, A. Acin, F. Verstraete, Discriminating states:
The quantum Chernoff bound, Phys. Rev. Lett. 98, 160501 (2007).

\bibitem{Maziero_NMuTP} J. Maziero, Non-monotonicity of trace distance
under tensor products, Braz. J. Phys. 45, 560 (2015). 

\bibitem{Lewenstein} S. Rana, P. Parashar, and M. Lewenstein, Trace-distance
measure of coherence, Phys. Rev. A 93, 012110 (2016). 

\bibitem{Li-1} L.-H. Shao, Z. Xi, H. Fan, and Y. Li, Fidelity and
trace-norm distances for quantifying coherence, Phys. Rev. A 91, 042120
(2015). 

\bibitem{Wang} Z. Wang, Y.-L. Wang, and Z.-X. Wang, Trace distance
measure of coherence for a class of qudit states, Quantum Inf. Process.
15, 4641 (2016). 

\bibitem{Kimura} G. Kimura, The Bloch vector for N-level systems,
Phys. Lett. A 314, 339 (2003). 

\bibitem{Krammer-1} R. A. Bertlmann and P. Krammer, Bloch vectors
for qudits, J. Phys. A: Math. Theor. 41, 235303 (2008).

\bibitem{Maziero_AMP} J. Maziero, Computing coherence vectors and
correlation matrices with application to quantum discord quantification,
Adv. Math. Phys. 2016, e6892178 (2016).

\bibitem{LibForQ} https://github.com/jonasmaziero/LibForQ.

\bibitem{Yang_RTC} A. Winter and D. Yang, Operational resource theory
of coherence, Phys. Rev. Lett. 116, 120404 (2016). 

\bibitem{Adesso_frozenC} T. R. Bromley, M. Cianciaruso, and G. Adesso,
Frozen quantum coherence, Phys. Rev. Lett., 114, 210401 (2015).

\bibitem{Guo_qutrit} J.-L. Guo, H. Li, and G.-L. Long, Decoherent
dynamics of quantum correlations in qubit\textendash{}qutrit systems,
Quantum Inf. Proc. 12, 3421 (2013).

\bibitem{Khan_qutrit} S. Khan and I. Ahmad, Environment generated
quantum correlations in bipartite qubit-qutrit systems, Optik 127,
2448 (2016).

\bibitem{Soares-Pinto2011} D. O. Soares-Pinto, M. H. Y. Moussa, J.
Maziero, E. R. deAzevedo, T. J. Bonagamba, R. M. Serra, and L. C.
C\'eleri, Equivalence between Redfield- and master-equation approaches
for a time-dependent quantum system and coherence control, Phys. Rev.
A 83, 062336 (2011). 

\bibitem{Maziero_LibForro} J. Maziero, Fortran code for generating
random probability vectors, unitaries, and quantum states, Front.
ICT 3, 4 (2016). 

\bibitem{LibForro} https://github.com/jonasmaziero/LibForro.\end{thebibliography}
\end{document}